%% file: main2.tex
\documentclass[a4paper,conference]{IEEEtran}
\IEEEoverridecommandlockouts
\usepackage{cite}
\usepackage{amsmath,amssymb,amsfonts}
\usepackage{algorithmic}
\usepackage{acronym}
\usepackage{graphicx}
\usepackage[caption=false, font=footnotesize, labelformat=simple]{subfig}
\usepackage[skip=2pt, belowskip=-6pt]{caption}

\usepackage{textcomp}
\usepackage{xcolor}
\def\BibTeX{{\rm B\kern-.05em{\sc i\kern-.025em b}\kern-.08em
    T\kern-.1667em\lower.7ex\hbox{E}\kern-.125emX}}
\usepackage{tikz}
\usepackage[font = footnotesize]{caption} 
\input{acronyms}

\input{math}
\newcommand{\bs}[1]{\boldsymbol{#1}}
\usepackage{booktabs}
\usepackage{microtype}
\usepackage{flushend}
\newcommand{\gpline}{\protect\tikz[baseline=-0.5ex]{\protect\draw[blue, thick] (0,0)--(0.4,0);}}
\newcommand{\objectcenter}{\protect\tikz[baseline=-0.5ex]{\protect\node[black, font=\bfseries] at (0,0) {+};}}
\newcommand{\measurepoints}{\protect\tikz[baseline=-0.5ex]{\protect\fill[green!100!lime] (0,0) circle (1.5pt);}}
\newcommand{\predinterval}{\protect\tikz[baseline=-0.5ex]{\protect\fill[blue!25] (0,0) rectangle (0.3,0.15);}}

\begin{document}
\bstctlcite{IEEEexample:BSTcontrol}   

\title{Gaussian Process-Based Extended Object Estimation for 6G ISAC at Millimeter-Wave Frequencies}

\author{
M. Ertug Pihtili\textsuperscript{\dag}, 
Ossi Kaltiokallio\textsuperscript{\dag}, 
Julia Equi\textsuperscript{\ddag}, 
Jukka Talvitie\textsuperscript{\dag}, 
Elena Simona Lohan\textsuperscript{\dag}, 
Ertugrul Basar\textsuperscript{\dag}, \\
and Mikko Valkama\textsuperscript{\dag} \\
\textsuperscript{\dag}Tampere Wireless Research Center, Electrical Engineering Unit, Tampere University, Tampere, Finland \\
\textsuperscript{\ddag}Ericsson Research, Jorvas, Finland \\
\vspace{-0.8cm}

}

\maketitle
\begin{abstract}
This paper introduces a \ac{gp}-based method for \ac{eoe} in integrated sensing and communication (ISAC) scenarios, representing a promising approach to enhance environmental awareness beyond the conventional point-scatterer assumption. The suitability of the proposed \ac{gp}-based method for \ac{eoe} is investigated through a practical measurement setup compliant with the fifth-generation (5G) New Radio (NR) standard and employing bistatic sensing, with results evaluated for both mapping and \ac{slam} cases at millimeter-wave (mmWave) frequencies. The findings reveal that the enhanced capabilities of communication networks, when combined with bistatic sensing and \ac{gp}-based \ac{eoe}, enable improved environmental awareness in future wireless systems. Importantly, the results demonstrate that, under practical conditions, \ac{gp} effectively performs \ac{eoe} in both mmWave mapping and \ac{slam} scenarios.
\end{abstract}

\acresetall
\section{Introduction}

{The \ac{6g} wireless systems will be capable of perceiving the environment by incorporating sensing into the network, in addition to providing enhanced communication functionalities \cite{Gonzalez2024}. In this context, \ac{ofdm} is a promising waveform, as it can serve both sensing and communication purposes. Notably, \ac{ofdm} combined with multiple-input multiple-output (MIMO) at millimeter-wave (mmWave) frequencies enables advanced sensing capabilities due to the increased bandwidth and larger array apertures, providing high resolution in both the delay and angular domains and allowing for resolvable multipath components. Consequently, the integrated sensing and communication (ISAC) paradigm will be a key enabler of functionalities such as localization, mapping, and \ac{slam}, thereby bringing enhanced environmental awareness to future wireless networks \cite{Gonzalez2024}.}

{In this context, conventional point-scatterer environmental object (or target) models assume that each multipath component originates from a single point object, referred to as a landmark in the environment. However, the high-resolution sensing capabilities of ISAC systems make it possible to move beyond this assumption by resolving multiple \acp{ip} originating from the same object, known as an extended object \cite{Zhai2025}. Consequently, non-line-of-sight (NLoS) paths can be exploited to enhance environmental awareness by sensing both active and passive objects, along with their physical characteristics such as shape, size, and material type \cite{Karttunen2025}.}

{An extended object refers to an object that generates multiple resolvable multipath rays when illuminated by a radio frequency (RF) signal, thereby enabling the estimation of its spatial extent. In \cite{Ge2025}, \ac{eoe} was performed by extracting the convex hull through a proposed post-processing scheme, which enhances landmark reconstruction and enables outlier removal for radio mapping, along with user equipment (UE) localization, within a bistatic sensing system framework. In \cite{Zhai2025}, a novel multipath-based \ac{slam} algorithm was proposed, incorporating the shape estimation of various object types through a constructed measurement model. Furthermore, \cite{Karstensen2025} introduced a multiple-model probability hypothesis density (PHD)-based \ac{slam} algorithm, which exploits both the measurements and the labels of the PHD filter for offline estimation of extended objects, such as reflecting surfaces and scattering objects, with results validated using practical measurements. Apart from environmental mapping and \ac{slam}, \cite{Baum2014} proposed a target tracking algorithm based on a random hypersurface model (RHM) for estimating star-convex shapes of extended objects, in addition to the commonly used ellipse assumptions. Additionally, \cite{Wahlstrom2015} introduced an extended target tracking scheme in which the shape of the extended target is estimated using \ac{gp}. Herein, \ac{gp} is a non-parametric Bayesian extension of linear regression, characterized by an infinitely long mean vector and an infinite-by-infinite covariance matrix, where any finite subset follows a joint Gaussian distribution \cite{Rasmussen2006}. While prior works have addressed \ac{eoe} in the ISAC literature, the experimental validation of \ac{gp}-based \ac{eoe} in practical ISAC scenarios remains unexplored, which motivates this work.}

{In this work, the suitability of GP in ISAC systems for \ac{eoe} using mmWave experimental data is investigated. The experiments are conducted using a fifth-generation (5G) new radio (NR) setup, where the \ac{tx} employs an OFDM waveform and transmits 5G NR standard-compliant beam-based positioning reference signals (PRSs) at 60~GHz. The experimental data are utilized for both mmWave mapping and \ac{slam} scenarios, in which the estimated \acp{ip} of single-bounce \ac{nlos} paths are subsequently used to approximate the contour of objects in the measurement environment by leveraging \acp{gp}, following the proposed method. Unlike the opportunistic case, where measurement points are assumed to originate from the entire contour of the object, the practical data capture only a portion of each object with a limited number of measurements. The findings indicate that, when combined with ISAC systems, \acp{gp} enable reliable estimation of object shapes even under limited measurement conditions, thereby enhancing environmental awareness.}

{\textit{Notation:} Italic lowercase letters $x$ denote scalars, bold lowercase letters $\mathbf{x}$ denote vectors, and bold uppercase letters $\mathbf{X}$ denote matrices. Complex- and real-valued vectors or matrices of size $a \times b$ are denoted by $\mathbb{C}^{a \times b}$ and $\mathbb{R}^{a \times b}$, respectively. The operator $(\cdot)^T$ represents transpose, and $\exp(x)$ is the exponential function. A real Gaussian distribution with zero mean and variance $\sigma^2$ is written as $\mathcal{N}(0, \sigma^2)$. The $N \times N$ identity matrix is denoted by $\mathbf{I}_N$, the norm by $\lVert \cdot \rVert$, and expectation by $\mathbb{E}[\cdot]$. Finally, $\mathbf{1}_N$ denotes an all-ones vector of size $N \times 1$.}

\section{Environmental Sensing}

{We consider a bistatic sensing scenario in which the \ac{tx} location $\vp_\textrm{TX} \in \mathbb{R}^2$ and orientation $\alpha_\textrm{TX}$ are known, the \ac{ip} of the $i$th propagation path is unknown $\vp_i \in \mathbb{R}^2$, and the \ac{rx} state is represented by $\vs_\textrm{RX} = [\vp_\textrm{RX}, \, \alpha_\textrm{RX}, \, b_\textrm{RX}]^\top \in \mathbb{R}^4$ which consists of the position, heading, and clock bias between the unsynchronized \ac{tx} and \ac{rx} clocks. In the following, we summarize two methods for estimating $\vp_i$'s using the channel parameter estimates $\vz_i$'s and the (un)known \ac{rx} state. The estimated \acp{ip} are then used by the GP-based \ac{eoe} method described in Section \ref{sec:extended_object_estimation}.}

\subsection{System Model}

We consider beam-based \ac{ofdm} transmission where beam sweeping is utilized at both \ac{tx} and \ac{rx}, equipped with antenna arrays. Assuming a time-invariant channel response over one \ac{ofdm} symbol, and sampling timing accuracy within the period of a cyclic prefix, the received sample of the $l$th OFDM-symbol and $k$th subcarrier for the $m$th \ac{tx} beam and $n$th \ac{rx} beam can be written as~\cite{rastorguevafoi2023}
\begin{equation} \label{eq:rx_symbol}
    y_{k,l}^{m,n}  =\sum_{i=0}^{N} \xi_i G^m_{\textrm{TX}}({\phi}_i) G^n_{\textrm{RX}}({\theta}_i) e^{-\imagunit2\pi k \Delta\!f \tau_i} x_{k,l}^{m,n} + w_{k,l}^{m,n}\!, 
\end{equation}
where $\xi_i$, $\tau_i$, ${\phi}_i$, ${\theta}_i$ are the complex path coefficient, \ac{toa}, \ac{aod}, and \ac{aoa} for the $i$th propagation path, respectively. Furthermore, $\Delta f$ denotes the subcarrier spacing, $x_{k,m} \in \mathbb{C}$ with $|x_{k,m}| = 1~\forall k,m$ is the transmitted PRS sample, and $w_{k,l}^{m,n}$ represents noise affected by the \ac{rx} beam. The total number of paths is $N+1$, where $N$ is the number of \ac{nlos} propagation paths, and the path index $i=0$ is reserved for the \ac{los} path, if present and detected. The effects of steering vectors and beamformers are captured by $G^{m}_{\textrm{TX}}({\phi}_i) \!\in\! \mathbb{C}$ and $G^{n}_{\textrm{RX}}({\theta}_i) \!\in\! \mathbb{C}$, which describe angular responses of the $m$th \ac{tx} beam and  $n$th \ac{rx} beam, respectively. Depending on the knowledge of $G^{m}_{\textrm{TX}}({\phi}_i)$ and $G^{n}_{\textrm{RX}}({\theta}_i)$, channel estimation can be implemented in multiple ways ~\cite{rastorguevafoi2023,shahmansoori2018}.

\subsection{mmWave Mapping}

Let $\vz_i = [\tau_i, \; \phi_i, \; \theta_i]^\top$ denote the channel parameter estimate of the $i$th propagation path. Assuming $\vz_i$ are unbiased and Gaussian, the likelihood function of the single-bounce propagation paths is \cite{kaltiokallio2025}:
\begin{equation}\label{eq:likelihood_function}
    p(\vz_i \mid \vs_\textrm{RX}, \vp_i) = \N(\vz_i \mid \vg(\vs_\textrm{RX}, \vp_i), \vR_i),
\end{equation}
where $\vR_i = \diag([1 \, \text{ns}, \, 1 \, \text{deg}, \, 1 \, \text{deg}]) \, \forall i$ is the covariance matrix and the mean is given by 
\begin{equation}
    \vg(\vs_\textrm{RX}, \vp_i) = \begin{bmatrix}
        \lVert \vp_\textrm{TX} - \vp_i \rVert/c + \lVert \vp_i - \vp_\textrm{RX} \rVert/c + b_\textrm{RX} \\
        \atan2(y_i - y_\textrm{TX},x_i - x_\textrm{TX}) - \alpha_\textrm{TX} \\
        \atan2(y_i - y_\textrm{RX},x_i - x_\textrm{RX}) - \alpha_\textrm{RX}
    \end{bmatrix},
\end{equation}
which describes the geometric relationship between the \ac{tx}, \ac{rx} and \ac{ip}. In the above, $c$ denotes the speed of light, and $\atan2(\cdot,\cdot)$ is the two-argument arctangent. Now, the \acp{ip} can be estimated independently for every propagation path, by solving the following nonlinear optimization problem
\begin{equation}\label{eq:mapping}
    \hat{\vp}_i = \underset{\vp_i}{\argmin} J(\vs_\textrm{RX}, \vp_i), \quad i = 1,\ldots,N, \vspace{-0.1cm}
\end{equation}
where $J(\vs_\textrm{RX}, \vp_i) = (\vz_i - \vg(\vs_\textrm{RX}, \vp_i))^\top \vR_i^{-1} (\vz_i - \vg(\vs_\textrm{RX}, \vp_i))$ is a quadratic cost function. It is to be noted that the described mapping method assumes $\vs_\textrm{RX}$ is known.

\subsection{mmWave Simultaneous Localization and Mapping}

\Ac{slam} aims to solve the unknown \ac{rx} state and \ac{ip} locations using a set of channel parameter estimates, $\cZ = \{\vz_0, \vz_1, \ldots, \vz_N \}$. The unknown parameters can be estimated if a sufficient number of single-bounce \ac{nlos} paths exist since each provides three additional known parameters (i.e., $\vz_i$), while being parameterized by two unknowns (i.e., $\vp_i$). To solve the \ac{slam} problem, we resort to a robust snapshot \ac{slam} algorithm described in \cite{kaltiokallio2025}. The algorithm employs a random sample consensus (RANSAC)-inspired \ac{ls} approach, in which the method first utilizes a minimal subset of $\cZ$ to compute an initial solution $\tilde{\vs}_\textrm{RX}$. Then, based on $\tilde{\vs}_\textrm{RX}$ and an error metric, $\cZ$ is partitioned into a set of inliers (i.e., \ac{los} and single-bounce \ac{nlos}) and outliers (i.e., multi-bounce \ac{nlos} and clutter). Thereafter, the problem is resolved using the inlier set and the final \ac{rx} estimate $\hat{\vs}_\textrm{RX}$ is the one that minimizes the \ac{ls} cost. Lastly, $\hat{\vp}_i$'s are computed independently for every inlier using \eqref{eq:mapping} and $\hat{\vs}_\textrm{RX}$. For further details, see \cite{kaltiokallio2025}.
\begin{figure*}[!t]
    \centering
    \subfloat[\label{fig: Rect_GP_func}]{\includegraphics[width=0.65\columnwidth]{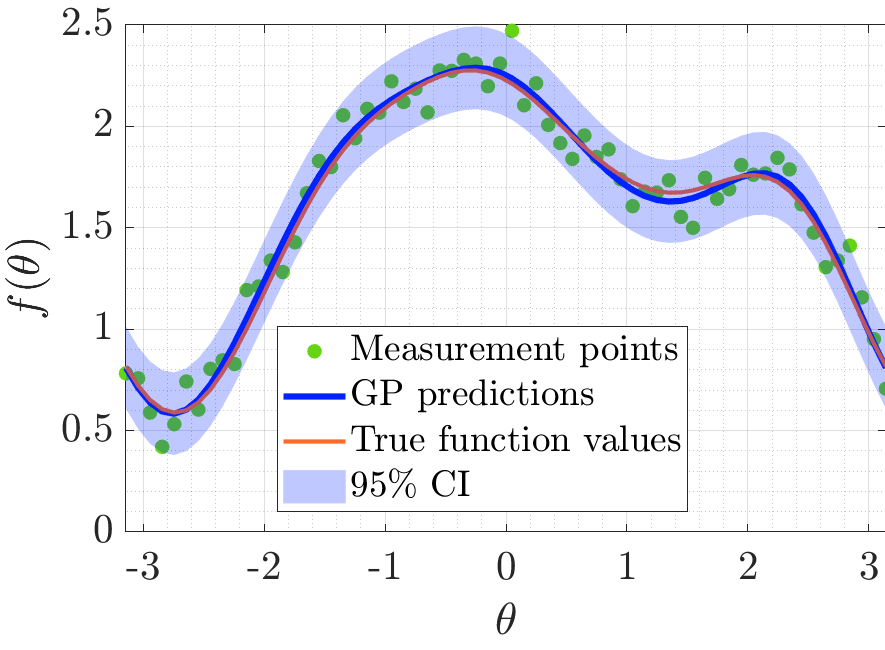}}
    \hfill
    \subfloat[\label{fig: Rect_GP_Est}]{\includegraphics[width=0.65\columnwidth]{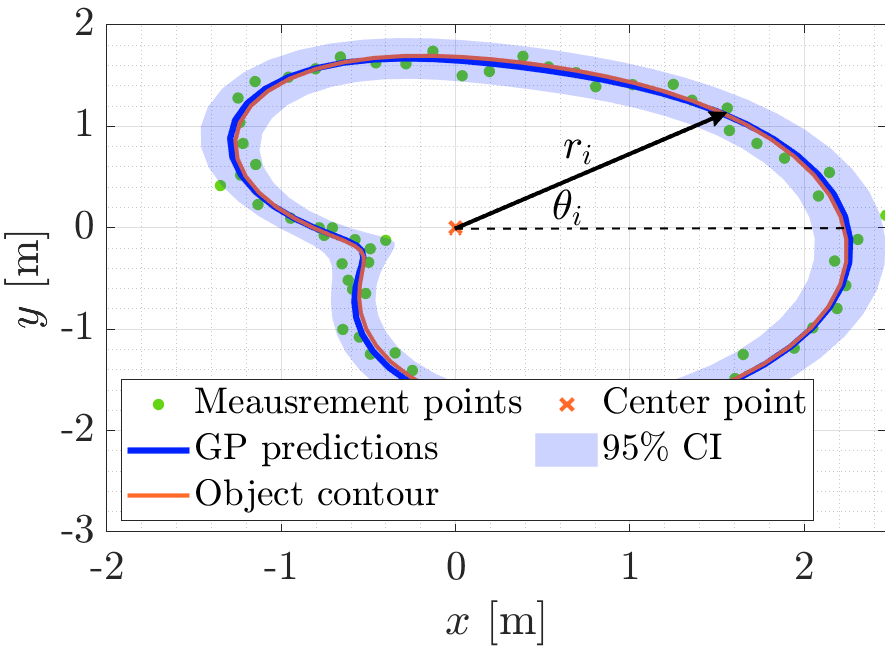}}
    \hfill
    \subfloat[\label{fig: BS_RMSE}]{\includegraphics[width=0.65\columnwidth]{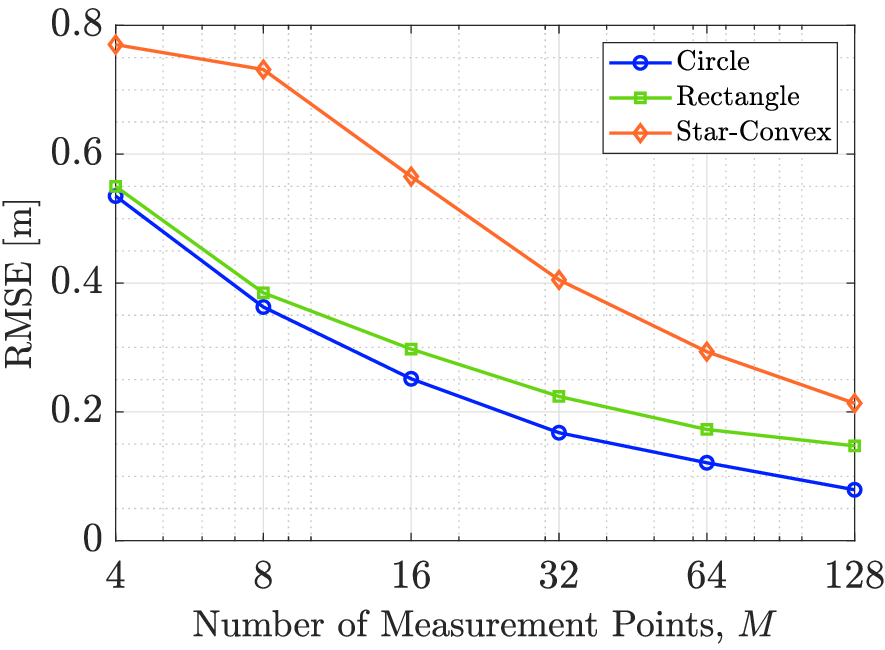}}
    \caption{Star-convex \ac{eoe} using \ac{gp} with $M = 64$ noisy contour measurements. The ground-truth parameters are $\bs{\beta} = [4.0, 4.0, 1.0, 0.01]$, while the estimated optimal values $\bs{\beta} = [0.9038, 0.3622, 1.6198, 0.0091]$ which are obtained by maximizing \eqref{eq: log_marg_lkh}. (a) Predicted function values with measurement points and the $95\%$ confidence interval (CI). (b) Reconstructed star-convex object showing both ground-truth and predicted contours. (c) RMSE performance for different extended objects versus the number of measurement points $M$.}
    \vspace{-0.3cm}
\end{figure*}

\section{GP-based Extended Object Estimation}

\subsection{Creating Training Dataset for GP-based EOE}\label{sec:clustering}

In order to perform GP-based EOE, the estimated IPs are first clustered by utilizing the density-based spatial clustering of applications with noise (DBSCAN) algorithm \cite{Ester1996}. Let $\hat{\vp}_i^c = [x_i^c, \, y_i^c]^\top, \; i \in \cI_c$ denote the cluster \acp{ip} which ideally capture the contour of the extended object and let $\cI_c$ denote the set of indices of cluster $c$ computed using DBSCAN. Since the \ac{gp}-based method describes the extended objects as star-convex shapes using a radial function \cite{Wahlstrom2015}, $\hat{\vp}_i^c$ are converted from the Euclidean coordinate system to the polar coordinate system as 
\begin{equation}
    \begin{bmatrix}
        \theta_i^c \\ r_i^c
    \end{bmatrix} = 
    \begin{bmatrix}
        \atan2 ( y_i^c - \bar{y}^c , x_i^c - \bar{x}^c ) \\ \sqrt{(x_i^c - \bar{x}^c)^2 + (y_i^c - \bar{y}^c)^2  }
    \end{bmatrix}, 
\end{equation}
where $\bar{x}^c = \tfrac{1}{M_c} \sum_{i \in \cI_c} x_i^c$ and $\bar{y}^c = \tfrac{1}{M_c} \sum_{i \in \cI_c} y_i^c$ denote the cluster centers, and $M_c = \lvert \cI_c \rvert$ cardinality of the set. In the next section, angles $\theta_i^c \in [-\pi,\pi]$ are used as training inputs (regressors) and radial distances $r_i^c$ are used as outputs (observations) of the GP regression model. Thus, collecting all regressors and observations in the cluster yields the training dataset used in \ac{gp}-based \ac{eoe}, mathematically given by
\begin{equation}
    \mathcal{D}_c = \Big\{\begin{bmatrix} \theta_i^c,\,  r_i^c \end{bmatrix}^\top \in \mathbb{R}^2 \,\Big|\, i \in \cI_c \Big\}.
\end{equation}

\subsection{Extended Object Estimation Using Gaussian Processes }\label{sec:extended_object_estimation}

In the following, \ac{gp}-based \ac{eoe} is performed independently for every cluster $\cD_c$ and dependence of cluster $c$ is excluded from the notation for brevity. Now considering dataset $\mathcal{D}$ which represents the contour of an extended object, the noisy radial observations can be defined as \cite{Wahlstrom2015}
\begin{equation}
    r_i = f(\theta_i) + n_i,
\end{equation}
where $f(\theta_i)$ is the radial function which we seek to estimate using \ac{gp} regression, and $n_i \sim \N(0, \sigma_n^2)$ denotes zero-mean \ac{iid} observation noise. Hence, the input–output mapping through the radial function $f(\theta_i)$ is modeled as a GP that maps polar angles to radial distances for two-dimensional (2D) object shape estimation, expressed as
\begin{equation}
    f(\theta_i) \sim \mathcal{GP}\left(m(\theta_i), \, k(\theta_i, \theta_{i'})\right), \label{eq:GP_nonzero}
\end{equation}
where the mean function $m(\theta_i)$ represents the average radius of the target contour and $k(\theta_i, \theta_{i'})$ is the covariance function which captures similarity between different function values. The GP is fully characterized by its mean, $\mathbb{E}[f(\theta_i)] = m(\theta_i)$, and its covariance function, 
$\mathbb{E}[\left(f(\theta_i)-m(\theta_i)\right) (f(\theta_{i'})-m(\theta_{i'}))] = k(\theta_i, \theta_{i'})$, where a periodic squared exponential covariance function with periodicity $2\pi$ is employed to capture the star-convex shape of the object, given by \cite{Wahlstrom2015}
\begin{equation}
   k(\theta_i, \theta_{i'})= \sigma_f^2 \exp\!\Bigg( -\frac{2 \sin^2\!\Big( \tfrac{\lVert \theta_i -  \theta_{i'} \rVert}{2} \Big)}{\ell^2} \Bigg). \label{eq: PSE_kernel}
\end{equation}
Therein, the covariance function quantifies the correlation among input points, and tuning its hyperparameters is essential for accurately modeling the input–output mapping. The hyperparameters in \eqref{eq: PSE_kernel} are as follows: $\sigma_f$ is the magnitude scale, which controls the amplitude of the function; and $\ell$ is the characteristic length scale, which determines the rate of variation of the function.

GP regression learns the values of $f(\theta_i)$ from the training inputs and predicts values for new test inputs $\theta_\ast$ by defining a distribution over functions. In principle, one could generate functions from the prior distribution $f(\theta_\ast) \sim \mathcal{N}(m(\theta_\ast), k(\theta_\ast, \theta_\ast))$ and discard those inconsistent with the data. However, this is not computationally feasible since there may be infinitely many such functions. Instead, the posterior distribution is obtained by combining the prior with the data, yielding \cite{Rasmussen2006}
\begin{equation}
    p\left(f(\theta_\ast)\,\big|\,\theta_\ast, \cD \right) 
    = \mathcal{N}\left(f(\theta_\ast) \,\big|\, \E[f(\theta_\ast)],\V[f(\theta_\ast)]\right). 
\end{equation}
As a result, the posterior leads to the key predictive equations of the GP, namely the predictive mean and predictive variance, which are given respectively as \cite{Rasmussen2006}
\begin{align}
        \E[f(\theta_\ast)] &= \mu_\ast + \vk_\ast^\top (\vK + \sigma_n^2 \vI_{M})^{-1} (\vr - \vmu), \label{eq: pred_mean} \\ 
        \V[f(\theta_\ast)] &= k(\theta_\ast, \theta_\ast) -  \vk_\ast^\top (\vK + \sigma_n^2 \vI_{M})^{-1} \vk_\ast.  \label{eq: pred_var}
\end{align}
In the above, $M = \lvert \cD \rvert$ is the number of training points in the cluster, $\vr = [r_1, \, r_2, \, \ldots, \, r_M]^\top$, $[\vK]_{ij} = k(\theta_i, \theta_j)$ and $\vk_\ast$ is an $M$-dimensional vector with the $i$th element being $[\vk_\ast]_{i} = k(\theta_\ast, \theta_i)$. The mean of the radial function $\mu$ is learned from data such that $\mu_\ast = \mu$ and $\vmu = \mathbf{1}_M \mu$.

In practice, to obtain an accurate estimate for $f(\bs{\theta}_\ast)$, the model parameters (mean $\mu$ and hyperparameters $\vbeta = [\sigma_f^2, \, \ell^2, \, \sigma_n^2]^\top$) should also be learned from data which can be achieved by maximizing the log marginal likelihood. The log marginal lihelihood is given by
\begin{equation}
     p(\mathbf{r} \mid \mu, \vbeta)
    = -\frac{1}{2}\bs{\eta}^T\mathbf{K}_y^{-1}\bs{\eta} 
    - \frac{1}{2} \log |\mathbf{K}_y| 
    - \frac{M}{2} \log 2\pi,
    \label{eq: log_marg_lkh}
\end{equation}
where $\bs{\eta} = \vr - \vmu$ and $\mathbf{K}_y = \vK + \sigma_n^2 \vI_{M}$. The partial derivatives of the log marginal likelihood are:
\begin{align}
    \frac{\partial \log p(\mathbf{r} \mid \mu, \bs{\beta})}{\partial \mu} 
    &= \mathbf{1}_M^\top \mathbf{K}_y^{-1}\bs{\eta}, \\
    \frac{\partial \log p(\mathbf{r} \mid \mu, \bs{\beta})}{\partial \beta_j} 
    &= \frac{1}{2}\text{tr}\!\left[(\bs{\alpha}\bs{\alpha}^T - \mathbf{K}_y^{-1})
    \frac{\partial \mathbf{K}_y}{\partial \beta_j}\right],
\end{align}
where $\bs{\alpha} = \mathbf{K}_y^{-1}\bs{\eta}$. Given the log marginal likelihood and its derivatives, the model parameters can now be trained, for example, using a gradient-based optimizer.

\subsection{GP-based EOE Example Using Synthetic Data}
{Figs.~\ref{fig: Rect_GP_func} and \ref{fig: Rect_GP_Est} illustrate \ac{gp}-based \ac{eoe} of a 2D star-convex shape based on $M = 64$ noisy radial observations which are generated synthetically. The \ac{gp} model parameters are trained by maximizing \eqref{eq: log_marg_lkh} and the predictive mean and variance are computed using \eqref{eq: pred_mean} and \eqref{eq: pred_var}, respectively. The estimated radial function and the $95\%$ confidence interval (CI), corresponding to the pointwise mean $\pm 2$ standard deviations, are shown in Fig.~\ref{fig: Rect_GP_func}.
Since the measurements cover the full object contour, the complete star-convex shape is predicted accurately as illustrated in Fig.~\ref{fig: Rect_GP_Est}. 

To demonstrate the effectiveness of \ac{gp}-based \ac{eoe}, Fig.~\ref{fig: BS_RMSE} shows the root mean square error (RMSE) between the true and predicted radial function values (in meters) for three basic extended objects under varying numbers of measurement points $M$, where the measurements are synthetically generated along the object contours. The RMSE is defined as $\text{RMSE} = \sqrt{\tfrac{1}{M} \sum_{i=1}^{M} \big(f(\theta_i) - \E[f(\theta_i)]\big)^2}$, and the results are averaged over $10^3$ Monte Carlo iterations. As observed, increasing the number of measurement points improves estimation performance in all cases, since more points provide more comprehensive information about the underlying input--output mapping. The results further demonstrate the robustness of \ac{gp}-based extended object estimation, as reliable approximations of the object contours are obtained even when the number of measurement points $M$ is small.}

\section{GP-based EOE Using mmWave Measurements}
\textls[-6]{In this section, the 2D \ac{eoe} using \acp{gp} with empirical mmWave measurements is presented for both the mapping and \ac{slam} scenarios. The measurements were conducted at the Hervanta Campus of Tampere University, Finland, inside the Campus Arena building, which features a large, partially open space with various environmental objects, including pillars and walls. The indoor mmWave measurements were obtained from 5G NR standard-compliant beam-based PRS transmissions with a bandwidth of $400\ \mathrm{MHz}$ \cite{3gpp38211}, where the \ac{tx} transmits an OFDM waveform at $60\ \mathrm{GHz}$, and the \ac{rx} follows a trajectory while collecting beamformed PRS signals for environmental sensing. Both the \ac{tx} and \ac{rx} are equipped with $4 \times 16$ uniform planar arrays (UPAs). More details on the measurement campaign and setup can be found in \cite{rastorguevafoi2023}.}

\begin{figure}[!t]
    \centering
    \subfloat[(a)][\label{fig: GP_mapping_Est}]{\includegraphics[width=0.90\columnwidth]{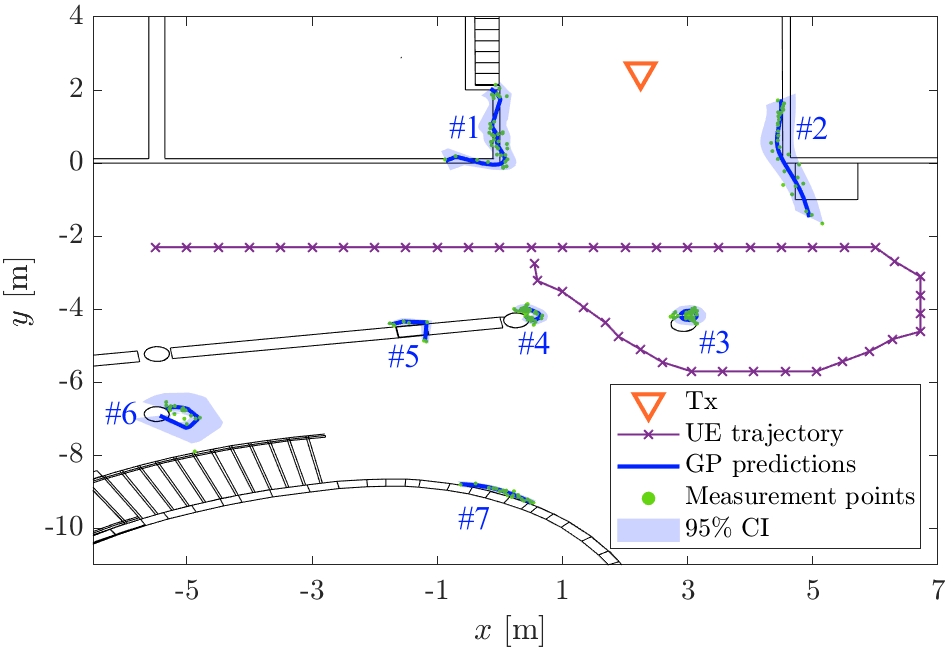}} \\ \vspace{-0.42cm}
    \subfloat[(b)][\label{fig: GP_SLAM_Est}]{\includegraphics[width=0.90\columnwidth]{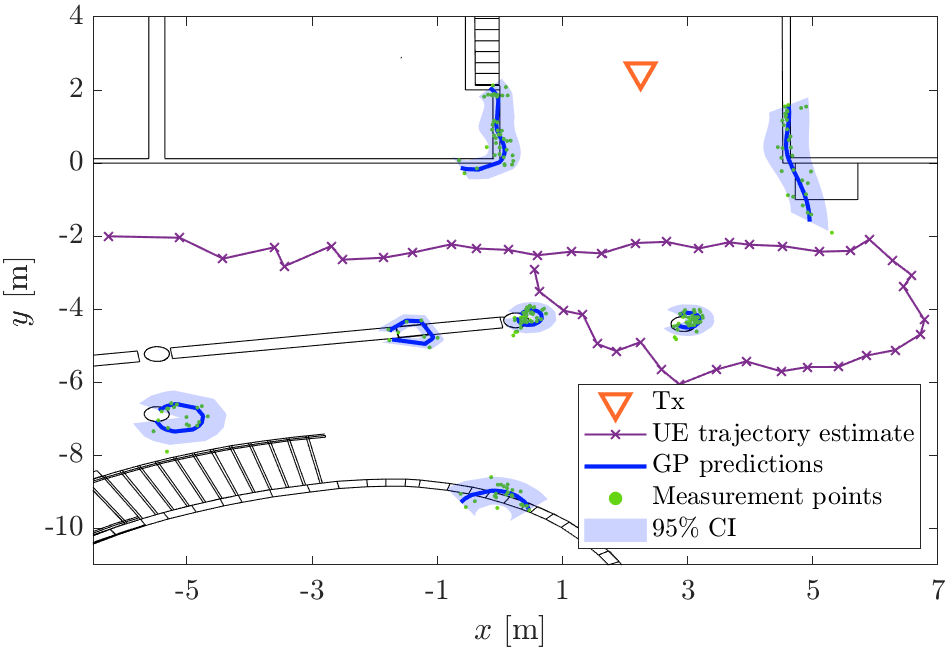}}
 \caption{(a) EOE using GP for mmWave mapping  scenario. (b) EOE using GP for mmWave \ac{slam}  scenario.}
\label{fig: GP_Est}
\end{figure}

\textls[-6]{Upon estimating the \acp{ip} either using mapping or \ac{slam}, the \acp{ip} are first clustered using the DBSCAN algorithm as described in Section \ref{sec:clustering}, resulting in seven clusters, each representing a set of measurement points corresponding to an object contour. Specifically, clusters 1, 2, 5, and 8 correspond to walls, while clusters 3, 4, and 6 correspond to pillars, as shown in Fig.~\ref{fig: GP_mapping_Est}. Subsequently, the created training datasets $\cD_c$ are used for \ac{gp}-based \ac{eoe} as presented in Section \ref{sec:extended_object_estimation}. For each cluster, the test inputs are selected as $\theta_i^\ast = \theta_i$, and the hyperparameters are initialized as $\ell = 2$, $\sigma_f = 2$, $\sigma_n = 2$, and $\mu = \mathbb{E}[\mathbf{r}]$ which denotes the mean radial distance of each cluster. The hyperparameters are then optimized by maximizing \eqref{eq: log_marg_lkh} to obtain the radial function $f(\boldsymbol{\theta})$ that best explains the input–output mapping. Furthermore, to better represent the wall geometries, biases are applied to the centroids of clusters~1,~2, and~7, defined as $[-1,\, 0]$, $[2,\, 0]$, and $[-0.2,\, -1]$, respectively.}
\begin{figure}[!t]
    \centering

    \subfloat[]{\includegraphics[width=0.49\columnwidth]{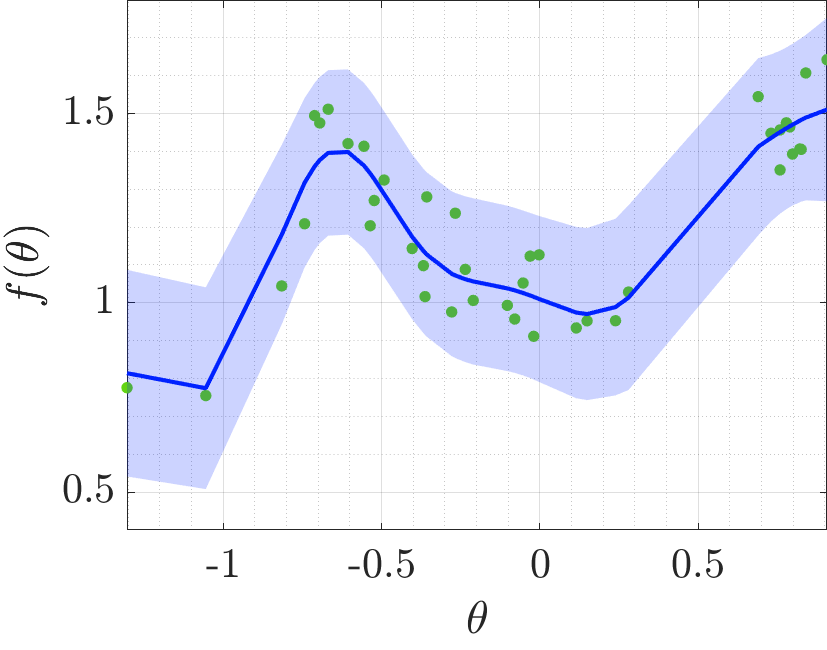}}
    \hfill
    \subfloat[]{\includegraphics[width=0.49\columnwidth]{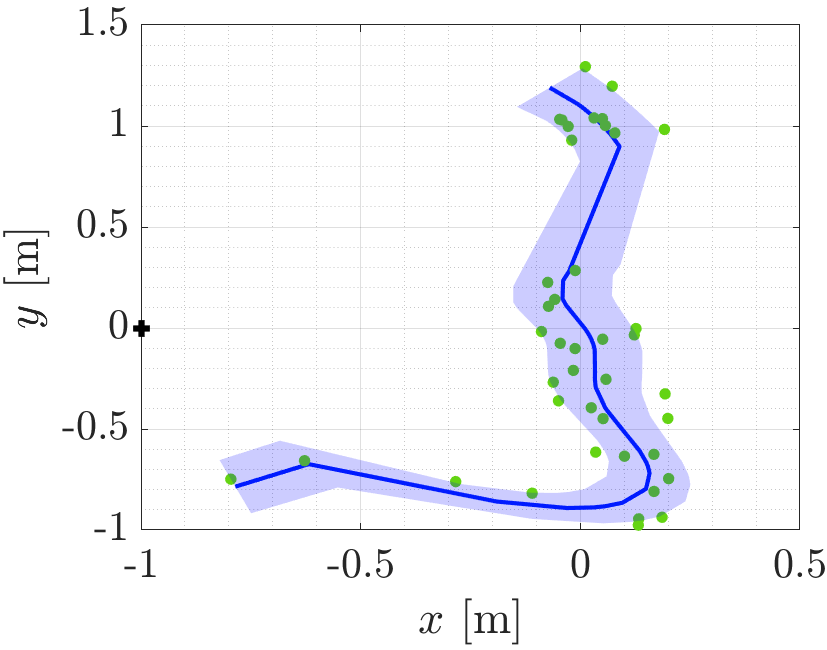}}
    \vspace{-0.38cm}
    \subfloat[]{\includegraphics[width=0.49\columnwidth]{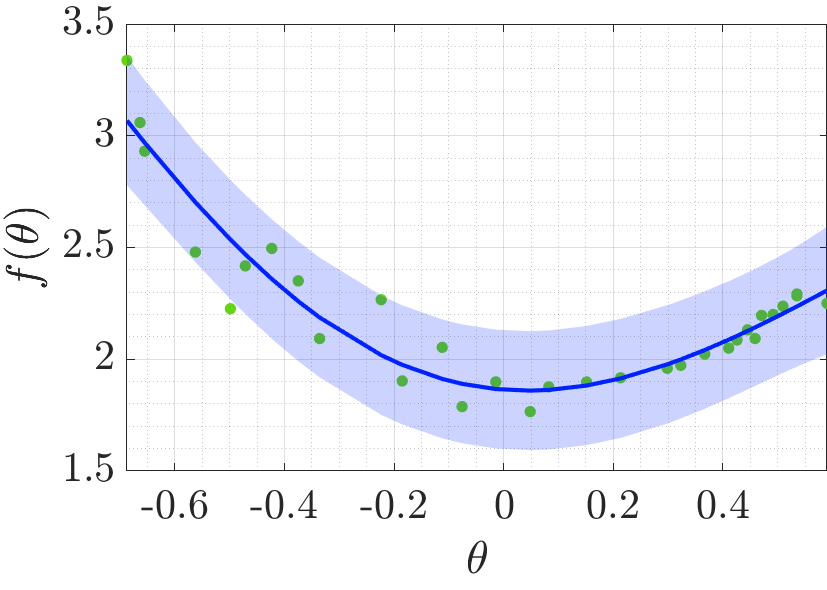}}
    \hfill
    \subfloat[]{\includegraphics[width=0.49\columnwidth]{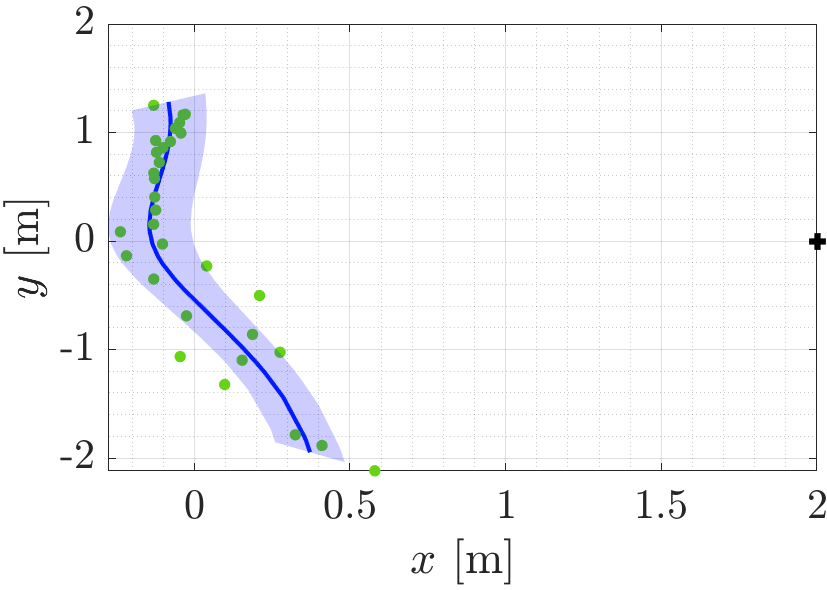}}
    \vspace{-0.38cm}
    \subfloat[]{\includegraphics[width=0.49\columnwidth]{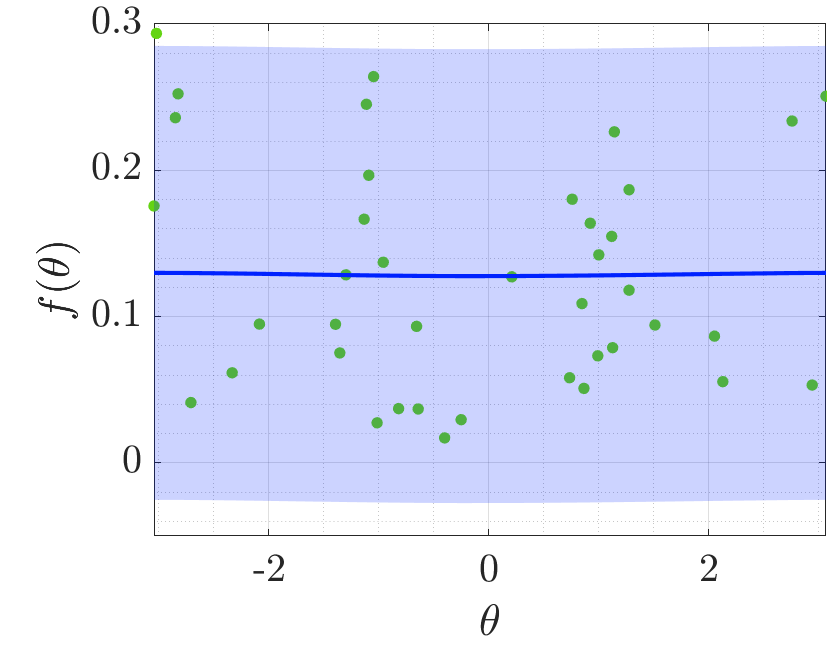}}
    \hfill
    \subfloat[]{\includegraphics[width=0.49\columnwidth]{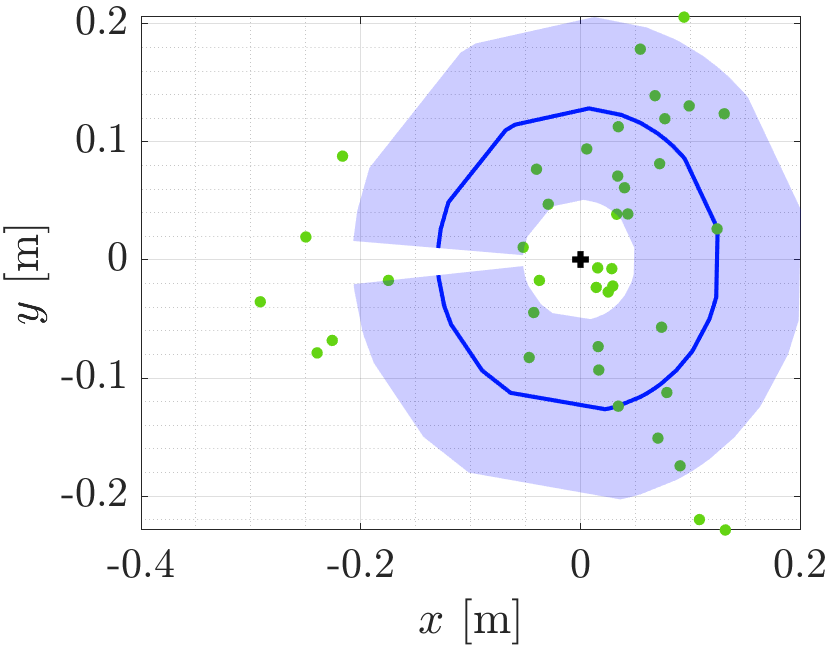}}
    \vspace{-0.38cm}
    \subfloat[]{\includegraphics[width=0.49\columnwidth]{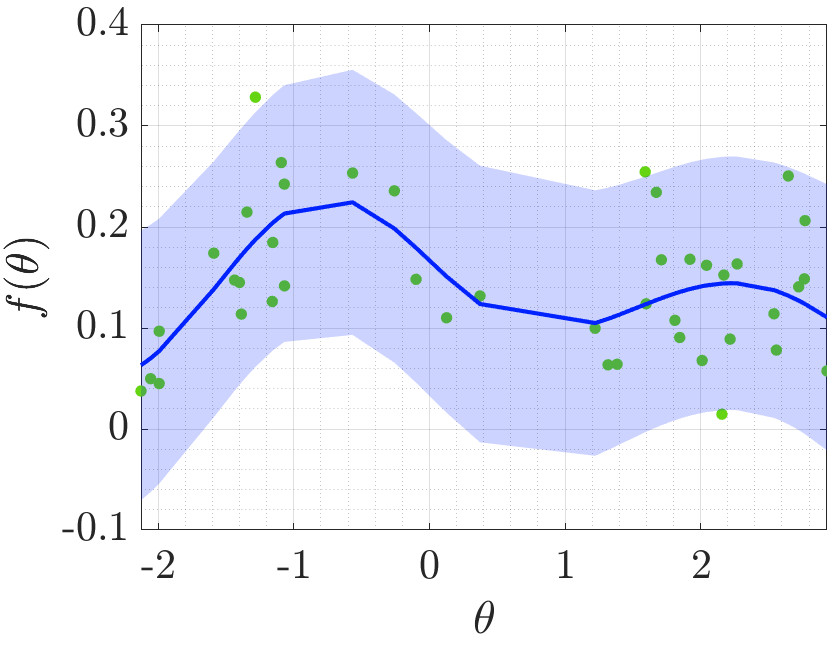}}
    \hfill
    \subfloat[]{\includegraphics[width=0.49\columnwidth]{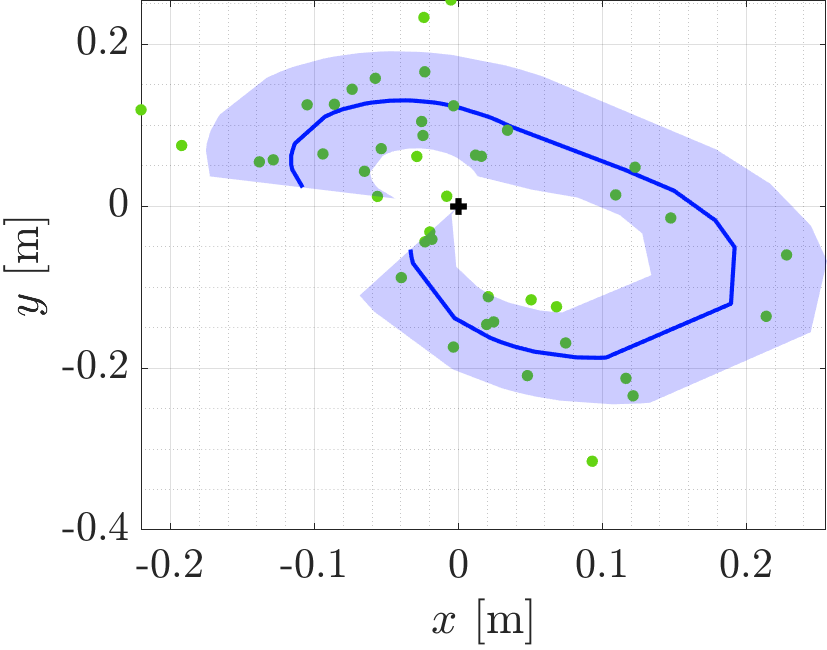}}
 \caption{(a), (c), (e), and (g) show the predictive mean functions in the radial domain, while (b), (d), (f), and (h) present the estimated object shapes in the Cartesian coordinate system for clusters 1, 2, 3, and 4, respectively. The legend indicates: \gpline~GP Predictions, \objectcenter~Object Center, \measurepoints~Measurement Points, \predinterval~$\%95$ Confidence Interval.  }
\label{fig: GP_mapping_func_Est}
\end{figure}

{Fig.~\ref{fig: GP_mapping_func_Est} presents the GP-based EOE results along with the predicted mean functions in \eqref{eq: pred_mean} for the mapping dataset, shown for clusters 1, 2, 3, and 4, respectively. In this context, $f(\theta)$ is a smooth function and ideally constant for circular shapes, meaning that test inputs enable the estimation of objects with smooth radial functions, such as clusters 3 and 4, even when only partial contour information is available. In contrast, objects with more complex geometries, such as walls, may exhibit rapidly varying radial functions $f(\theta)$, making estimation more challenging under limited data, as observed for clusters 1 and 2. For instance, the optimized characteristic length-scale values of cluster~1 (a wall) and cluster~3 (a pillar) are $\ell_1 = 0.2343$ and $\ell_3 = 1.3129$ for the mapping dataset. A smaller $\ell$ indicates faster variations in $f({\theta})$, as $\ell$ determines the distance over which the function changes significantly.}

{Ultimately, Fig.~\ref{fig: GP_Est} illustrates the GP prediction results for the empirical mmWave measurements, together with the map of the sensing environment. Comparing Figs.~\ref{fig: GP_mapping_Est} and \ref{fig: GP_SLAM_Est}, in the mapping scenario, the \ac{rx} position and ground-truth trajectory are known, whereas in the \ac{slam} scenario, the \ac{rx} location is unknown and its trajectory is estimated. Although the \ac{ip} estimates in the \ac{slam} scenario are less accurate than those in the mapping case, the overall performance of both scenarios remains comparable, indicating that the GP-based EOE performs well in both settings. It is important to emphasize that the predictive capability of GP is inherently constrained by both the number of measurement points and the angular coverage of the sampled object contours, since in both scenarios, the measurement points do not cover the entire object contour. Consequently, the training inputs, i.e., the polar angles ${\theta}_i$, represent only the portions of the objects that were sampled, which reflects a practical scenario as capturing the full contour with a single \ac{tx}–\ac{rx} pair is often infeasible. Even with partial information, the shapes of the pillars (clusters~3,~4, and~6), which correspond to circles in 2D, can still be accurately predicted. The GP model provides only coarse estimates of the wall extents based on the available measurement points, indicating that clusters~1 and~2 require additional measurements spanning a wider angular range of the object contour for improved EOE performance, a limitation that can be mitigated by employing multiple \ac{tx}–\ac{rx} pairs to capture the full object contours. }

\section{Conclusion}
{This paper has presented a GP-based EOE approach using practical mmWave data, where the measurement setup follows 5G NR standards. The results demonstrate that object shapes in the environment can be either finely or coarsely estimated depending on the object geometry and the availability of measurement data along the object contour. When combined with the environmental sensing capabilities of ISAC systems, GP provides valuable information about the surrounding environment, and the estimated extended objects can be further leveraged for ISAC applications by offering more concrete structural insights compared to conventional methods that rely on point scatterer assumptions. This information can be used for environmental mapping and for identifying static objects along with their shapes, which can subsequently enhance clutter cancellation algorithms in ISAC systems, leading to more accurate sensing performance than approaches based on point-scatterer models.}

\section*{Acknowledgement}
This work was funded by the EU Horizon Europe MSCA (MiFuture, Grant No. 101119643) on ultra-massive MIMO for future cell-free heterogeneous networks.
\bibliographystyle{IEEEtran}
\bibliography{ref}

\end{document}

%% file: acronyms.tex
\acrodef{eoe}[EOE]{extended object estimation}
\acrodef{gp}[GP]{Gaussian process}
\acrodef{tx}[Tx]{transmitter}
\acrodef{rx}[Rx]{receiver}
\acrodef{bs}[BS]{base station}
\acrodef{ue}[UE]{user equipment}
\acrodef{toa}[ToA]{time-of-arrival}
\acrodef{aod}[AoD]{angle-of-departure}
\acrodef{aoa}[AoA]{angle-of-arrival}
\acrodef{ip}[IP]{incidence point}
\acrodef{los}[LoS]{line-of-sight}
\acrodef{nlos}[NLoS]{non-line-of-sight}
\acrodef{ofdm}[OFDM]{orthogonal frequency-division multiplexing}
\acrodef{ls}[LS]{least squares}
\acrodef{ransac}[RANSAC]{random sample consensus}
\acrodef{slam}[SLAM]{simultaneous localization and mapping}
\acrodef{6g}[6G]{sixth-generation}
\acrodef{iid}[i.i.d.]{independent and identically distributed }

%% file: math.tex
\newcommand{\N}[0]{\ensuremath{\mathcal{N}}}
\newcommand{\cZ}[0]{\ensuremath{\mathcal{Z}}}
\newcommand{\cI}[0]{\ensuremath{\mathcal{I}}}
\newcommand{\cD}[0]{\ensuremath{\mathcal{D}}}
\newcommand{\E}[0]{\ensuremath{\mathbb{E}}}
\newcommand{\V}[0]{\ensuremath{\mathbb{V}}}
\newcommand{\imagunit}{\mathrm{j}}
\DeclareMathOperator{\diag}{diag}
\DeclareMathOperator{\atan2}{atan2}
\DeclareMathOperator{\argmin}{arg \, min \,}

\renewcommand{\vec}[1]{\ensuremath{{\mathbf{#1}}}}
\newcommand{\vp}[0]{\vec{p}}
\newcommand{\vs}[0]{\vec{s}}

\newcommand{\vz}[0]{\vec{z}}
\newcommand{\vg}[0]{\vec{g}}
\newcommand{\vR}[0]{\vec{R}}
\newcommand{\vk}[0]{\vec{k}}
\newcommand{\vK}[0]{\vec{K}}
\newcommand{\vI}[0]{\vec{I}}
\newcommand{\vr}[0]{\vec{r}}

\newcommand{\vecsymbol}[1]{\ensuremath{\boldsymbol{#1}}}

\newcommand{\vbeta}[0]{\vecsymbol{\beta}}
\newcommand{\vmu}[0]{\vecsymbol{\mu}}

%% file: ref.bib
@IEEEtranBSTCTL{IEEEexample:BSTcontrol,
CTLuse_forced_etal       = "yes",
CTLmax_names_forced_etal = "4",
CTLnames_show_etal       = "1" }

@ARTICLE{rastorguevafoi2023,
  author={Rastorgueva-Foi, Elizaveta and Kaltiokallio, Ossi and Ge, Yu and Turunen, Matias and Talvitie, Jukka and Tan, Bo and Furkan Keskin, Musa and Wymeersch, Henk and Valkama, Mikko},
  journal={IEEE J. Sel. Areas Commun.}, 
  title={Millimeter-wave Radio {SLAM}: End-to-End Processing Methods and Experimental Validation}, 
  year={2024},
  volume={42},
  number={9},
  pages={2550-2567},
  doi={10.1109/JSAC.2024.3413995}
}

@ARTICLE{shahmansoori2018,  
author={Shahmansoori, Arash and Garcia, Gabriel E. and Destino, Giuseppe and Seco-Granados, Gonzalo and Wymeersch, Henk},  
journal={IEEE Trans. Wireless Commun.},  
title={Position and Orientation Estimation Through Millimeter-Wave {MIMO} in {5G} Systems}, 
year={2018},  
volume={17}, 
number={3},  
pages={1822-1835},  
doi={10.1109/TWC.2017.2785788}
}

@ARTICLE{kaltiokallio2025,
  author={Kaltiokallio, Ossi and Rastorgueva-Foi, Elizaveta and Talvitie, Jukka and Ge, Yu and Wymeersch, Henk and Valkama, Mikko},
  journal={IEEE Trans. Veh. Technol.}, 
  title={Robust Snapshot Radio {SLAM}}, 
  year={2025},
  volume={74},
  number={5},
  pages={8460-8465},
  doi={10.1109/TVT.2024.3524118}
}

@ARTICLE{Wahlstrom2015,
  author={Wahlström, Niklas and Özkan, Emre},
  journal={IEEE Trans. Signal Process.}, 
  title={Extended Target Tracking Using {Gaussian} Processes}, 
  year={2015},
  volume={63},
  number={16},
  pages={4165-4178},
  doi={10.1109/TSP.2015.2424194}
}

@book{Rasmussen2006,
  author = {Rasmussen, Carl Edward and Williams, Christopher K. I.},
  publisher = {The MIT Press},
  title = {Gaussian Processes for Machine Learning},
  year = 2006
}

@standard{3gpp38211,
  title        = {NR; Physical Channels and Modulation},
  organization = {3GPP},
  number       = {TS 38.211},
  edition      = {Release 18},
  year         = {2023},
  month        = sep,
}

@ARTICLE{Gonzalez2024,
  author={González-Prelcic, Nuria and Furkan Keskin, Musa and Kaltiokallio, Ossi and Valkama, Mikko and Dardari, Davide and Shen, Xiao and Shen, Yuan and Bayraktar, Murat and Wymeersch, Henk},
  journal={Proc. IEEE}, 
  title={The Integrated Sensing and Communication Revolution for {6G}: Vision, Techniques, and Applications}, 
  year={2024},
  volume={112},
  number={7},
  pages={676-723},
  doi={10.1109/JPROC.2024.3397609}
}

@misc{Ge2025,
 title={Pilot-Based End-to-End Radio Positioning and Mapping for {ISAC}: Beyond Point-Based Landmarks}, 
 author={Yu Ge and Musa Furkan Keskin and Hui Chen and Ossi Kaltiokallio and Mengting Li and Mikko Valkama and Christos Masouros and Henk Wymeersch},
 year={2025},
 eprint={2505.07402},
 archivePrefix={arXiv},
 primaryClass={eess.SP},
 url={https://arxiv.org/abs/2505.07402}, 
}

@ARTICLE{Zhai2025,
  author={Zhai, Shiyu and Fan, Jiancun and Gao, Jiawei and Dai, Gang},
  journal={IEEE Trans. Wireless Commun.}, 
  title={Multipath-Based {SLAM} Exploiting Extended Object Estimation and Classification}, 
  year={2025},
  volume={24},
  number={8},
  pages={7029-7045},
  doi={10.1109/TWC.2025.3557580}
}

@INPROCEEDINGS{Karttunen2025,
  author={Karttunen, Aki and Talvitie, Jukka and Kaltiokallio, Ossi and Rastorgueva-Foi, Elizaveta and Valkama, Mikko},
  booktitle={2025 IEEE 5th Int. Symp. Joint Commun. \& Sensing (JC\&S)}, 
  title={Towards Semantic Radio {SLAM} with Landmark Feature Extraction in mmWave Networks}, 
  year={2025},
  pages={1-6},
  doi={10.1109/JCS64661.2025.10880651}
}

@INPROCEEDINGS{Karstensen2025,
  author={Karstensen, Peter Iwer Hoedt and Kaltiokallio, Ossi and Rastorgueva-Foi, Elizaveta and Talvitie, Jukka and Valkama, Mikko},
  booktitle={2025 19th Eur. Conf. Antennas Propag. (EuCAP)}, 
  title={Bistatic Radio {SLAM} with Offline Shape Estimation}, 
  year={2025},
  pages={1-5},
  doi={10.23919/EuCAP63536.2025.11000082}
}

@ARTICLE{Baum2014,
  author={Baum, Marcus and Hanebeck, Uwe D.},
  journal={IEEE Trans. Aerosp. Electron. Syst.}, 
  title={Extended Object Tracking with Random Hypersurface Models}, 
  year={2014},
  volume={50},
  number={1},
  pages={149-159},
  doi={10.1109/TAES.2013.120107}
}

@inproceedings{Ester1996,
  title={A density-based algorithm for discovering clusters in large spatial databases with noise},
  author={Ester, Martin and Kriegel, Hans-Peter and Sander, J{\"o}rg and Xu, Xiaowei and others},
  booktitle={kdd},
  volume={96},
  number={34},
  pages={226--231},
  year={1996}
}
